\documentclass[12pt]{article}

\usepackage [latin1]{inputenc}

\usepackage{lscape}
\usepackage{graphics}
\usepackage{epsfig}
\usepackage{graphicx}
\usepackage{color}
\usepackage{deluxetable} 
\usepackage{amsmath, amsthm} 

\begin{document}
\baselineskip=24pt

\newcommand{\D}{\displaystyle} 
\newcommand{\T}{\textstyle} 
\newcommand{\SC}{\scriptstyle} 
\newcommand{\SSC}{\scriptscriptstyle} 

\newcommand{\be}{\begin{eqnarray}}
\newcommand{\ee}{\end{eqnarray}}

\definecolor{yellow}{rgb}{0.95,0.75,0.1}
\definecolor{orange}{rgb}{0.95,0.4,0.1}
\definecolor{red}{rgb}{1,0,0}
\definecolor{green}{rgb}{0,1,0}
\definecolor{blue}{rgb}{0,0.5,1}

\definecolor{lblue}{rgb}{0,0.8,1}
\definecolor{dblue}{rgb}{0,0,1}
\definecolor{dgreen}{rgb}{0,0.65,0}
\definecolor{lila}{rgb}{0.8,0,0.8}
\definecolor{violet}{rgb}{1,0,0.9}
\definecolor{grey}{rgb}{0.3,0.3,0.3}

\definecolor{contoura}{rgb}{0,0,1}
\definecolor{contourb}{rgb}{0,1,1}
\definecolor{contourc}{rgb}{0,1,0}
\definecolor{contourd}{rgb}{0.95,0.75,0.1}
\definecolor{contoure}{rgb}{1,0,0}
\definecolor{contourf}{rgb}{1,0,1}

\newcommand\cred[1]{\textcolor{red}{#1}}
\newcommand\cblue[1]{\textcolor{blue}{#1}}
\newcommand\ccyan[1]{\textcolor{contourb}{#1}}
\newcommand\cgreen[1]{\textcolor{green}{#1}}
\newcommand\cyellow[1]{\textcolor{yellow}{#1}}
\newcommand\cviolet[1]{\textcolor{violet}{#1}}
\newcommand\cmagenta[1]{\textcolor{magenta}{#1}}
\newcommand\cgrey[1]{\textcolor{grey}{#1}}

\newcommand\cconta[1]{\textcolor{contoura}{#1}}
\newcommand\ccontb[1]{\textcolor{contourb}{#1}}
\newcommand\ccontc[1]{\textcolor{contourc}{#1}}
\newcommand\ccontd[1]{\textcolor{contourd}{#1}}
\newcommand\cconte[1]{\textcolor{contoure}{#1}}
\newcommand\ccontf[1]{\textcolor{contourf}{#1}}

\newcommand {\black} {\color{black}}
\newcommand {\violet}{\color{violet}}
\newcommand {\blue} {\color{blue}}
\newcommand {\dblue} {\color{dblue}}
\newcommand {\cyan} {\color{cyan}}
\newcommand {\lila} {\color{lila}}
\newcommand {\yellow} {\color{yellow}}
\newcommand {\green} {\color{green}}
\newcommand {\dgreen} {\color{dgreen}}
\newcommand {\magenta} {\color{magenta}}
\newcommand {\red} {\color{red}}
\newcommand {\orange} {\color{orange}}

\def\AJ{{\it Astron. J.} }
\def\ARAA{{\it Annual Rev. of Astron. \& Astrophys.} }
\def\ARNPS{{\it Annual Rev. of Nucl. \& Part. Sci. } }
\def\ApJ{{\it Astrophys. J.} }
\def\ApJL{{\it Astrophys. J. Letters} }
\def\ApJS{{\it Astrophys. J. Suppl.} }
\def\ApP{{\it Astropart. Phys.} }
\def\AA{{\it Astron. \& Astroph.} }
\def\AAR{{\it Astron. \& Astroph. Rev.} }
\def\AAL{{\it Astron. \& Astroph. Letters} }
\def\AASu{{\it Astron. \& Astroph. Suppl.} }
\def\AN{{\it Astron. Nachr.} }
\def\ASR{{\it Adv. in Space Res.} }
\def\EPJC{{\it Eur. Phys. Journ.} {\bf C} }
\def\IJMP{{\it Int. J. of Mod. Phys.} }
\def\JCAP{{\it J. of Cosmol. and Astrop. Phys.} }
\def\JGR{{\it Journ. of Geophys. Res.}}
\def\JHEP{{\it Journ. of High En. Phys.} }
\def\JPhG{{\it Journ. of Physics} {\bf G} }
\def\CQG{{\it Class. Quant. Grav. } }
\def\MNRAS{{\it Month. Not. Roy. Astr. Soc.} }
\def\Nature{{\it Nature} }
\def\NewAR{{\it New Astron. Rev.} }
\def\NewA{{ New Astron.} }
\def\NIMPA{{\it Nucl. Instr. Meth. Phys. Res.}{\bf A} }
\def\PASP{{\it Publ. Astron. Soc. Pac.} }
\def\PhFl{{\it Phys. of Fluids} }
\def\PLB{{\it Phys. Lett.}{\bf B} }
\def\PR{{\it Phys. Rev.} }
\def\PRD{{\it Phys. Rev.} {\bf D} }
\def\PRL{{\it Phys. Rev. Letters} }
\def\PRX{{\it Phys. Rev. }{\bf X} }
\def\RMP{{\it Rev. Mod. Phys.} }
\def\RPP{{\it Rep. Pro.Phys.} }
\def\Science{{\it Science} }
\def\ZfA{{\it Zeitschr. f{\"u}r Astrophys.} }
\def\ZfN{{\it Zeitschr. f{\"u}r Naturforsch.} }
\def\etal{{\it et al.}}

\hyphenation{mono-chro-matic  sour-ces  Wein-berg
chang-es Strah-lung dis-tri-bu-tion com-po-si-tion elec-tro-mag-ne-tic
ex-tra-galactic ap-prox-i-ma-tion nu-cle-o-syn-the-sis re-spec-tive-ly
su-per-nova su-per-novae su-per-nova-shocks con-vec-tive down-wards
es-ti-ma-ted frag-ments grav-i-ta-tion-al-ly el-e-ments me-di-um
ob-ser-va-tions tur-bul-ence sec-ond-ary in-ter-action
in-ter-stellar spall-ation ar-gu-ment de-pen-dence sig-nif-i-cant-ly
in-flu-enc-ed par-ti-cle sim-plic-i-ty nu-cle-ar smash-es iso-topes
in-ject-ed in-di-vid-u-al nor-mal-iza-tion lon-ger con-stant
sta-tion-ary sta-tion-ar-i-ty spec-trum pro-por-tion-al cos-mic
re-turn ob-ser-va-tion-al es-ti-mate switch-over grav-i-ta-tion-al
super-galactic com-po-nent com-po-nents prob-a-bly cos-mo-log-ical-ly
Kron-berg Berk-huij-sen}
\def\simle{\lower 2pt \hbox {$\buildrel < \over {\scriptstyle \sim }$}}
\def\simge{\lower 2pt \hbox {$\buildrel > \over {\scriptstyle \sim }$}}
\def\intunits{{\rm s}^{-1}\,{\rm sr}^{-1} {\rm cm}^{-2}}

\def\sun{\hbox{$\odot$}}
\def\la{\mathrel{\mathchoice {\vcenter{\offinterlineskip\halign{\hfil
$\displaystyle##$\hfil\cr<\cr\sim\cr}}}
{\vcenter{\offinterlineskip\halign{\hfil$\textstyle##$\hfil\cr
<\cr\sim\cr}}}
{\vcenter{\offinterlineskip\halign{\hfil$\scriptstyle##$\hfil\cr
<\cr\sim\cr}}}
{\vcenter{\offinterlineskip\halign{\hfil$\scriptscriptstyle##$\hfil\cr
<\cr\sim\cr}}}}}
\def\ga{\mathrel{\mathchoice {\vcenter{\offinterlineskip\halign{\hfil
$\displaystyle##$\hfil\cr>\cr\sim\cr}}}
{\vcenter{\offinterlineskip\halign{\hfil$\textstyle##$\hfil\cr
>\cr\sim\cr}}}
{\vcenter{\offinterlineskip\halign{\hfil$\scriptstyle##$\hfil\cr
>\cr\sim\cr}}}
{\vcenter{\offinterlineskip\halign{\hfil$\scriptscriptstyle##$\hfil\cr
>\cr\sim\cr}}}}}
\def\degr{\hbox{$^\circ$}}
\def\arcmin{\hbox{$^\prime$}}
\def\arcsec{\hbox{$^{\prime\prime}$}}
\def\utw{\smash{\rlap{\lower5pt\hbox{$\sim$}}}}
\def\udtw{\smash{\rlap{\lower6pt\hbox{$\approx$}}}}
\def\fd{\hbox{$.\!\!^{\rm d}$}}
\def\fh{\hbox{$.\!\!^{\rm h}$}}
\def\fm{\hbox{$.\!\!^{\rm m}$}}
\def\fs{\hbox{$.\!\!^{\rm s}$}}
\def\fdg{\hbox{$.\!\!^\circ$}}
\def\farcm{\hbox{$.\mkern-4mu^\prime$}}
\def\farcs{\hbox{$.\!\!^{\prime\prime}$}}
\def\fp{\hbox{$.\!\!^{\scriptscriptstyle\rm p}$}}
\def\baselinestretch{1.5}
\def\gsim{\stackrel{>}{\sim}}
\def\lsim{\stackrel{<}{\sim}}
\def\beq{\begin{equation}}
\def\eeq{\end{equation}}
\def\ol{\overline}


\centerline{Frontiers in Astronomy}
%
%
\vskip0.5cm
\begin{centering}
{A two-step strategy to identify episodic sources of gravitational waves and high energy neutrinos in starburst galaxies}%
\footnote[1]{Corresponding author: Peter L. Biermann, plbiermann@mpifr-bonn.mpg.de. {\bf \boldmath $^{\dagger}$} refers to coauthors who are no longer among us}



{  M. Allen} (Department of Physics \& Astronomy, Washington State University, Pullman, WA 99164, USA); 
{  P.L. Biermann} (i) MPI for Radioastr., Auf dem H{\"u}gel 69, D-53121 Bonn, Germany; ii) Dept. of Phys. \& Astron., U. Alabama, Box 870324, Tuscaloosa, AL 35487-0324, USA); 
{\bf   L.I. Caramete} (Cosmology and AstroParticle Physics Group, Institute of Space Science, Bucharest-Magurele, Romania);  
{  A. Chieffi} (Istituto Nazionale Di Astrofisica (INAF) $-$ Istituto di Astrofisica e Planetologia Spaziali, Via Fosso del Cavaliere 100, I-00133 Roma, Italy);  
{  R. Chini} (i) Astronomisches Institut, Ruhr$-$Universit\"at Bochum, Universit\"atsstra{\ss}e 150, D-44801 Bochum, Germany; ii) Centrum Astronomiczne im. Mikolaja Kopernika, PAN, Bartycka 18, PL-00-716 Warsaw, Poland; iii) Instituto de Astronom\'{i}a, Universidad Cat\'{o}lica del Norte, Avenida Angamos 0610, Antofagasta, Chile); 
{  D. Frekers} (Institut f{\"u}r Kernphysik, Westf{\"a}lische Wilhelms-Universit{\"a}t M{\"u}nster, D-48149 M{\"u}nster, Germany); 
{  L. Gergely} (i) Univ. Szeged, Department of Theoretical Physics, Szeged, Hungary; ii) HUN-REN (Hungarian Research Network) Wigner Research Centre for Physics, Department of Theoretical Physics, Budapest, Hungary); 
{  B. Harms$^{\dagger}$} (Dept. of Phys. \& Astron., U. Alabama, Box 870324, Tuscaloosa, AL 35487-0324, USA);  
{  I. Jaroschewski} (Faculty of Physics and Astronomy, Ruhr$-$Universit\"at Bochum, Universit\"atsstra{\ss}e 150, D-44801 Bochum, Germany); %
{  P.S. Joshi} (Int. Center for Space \& Cosmology, Ahmedabad Univ., Ahmedabad 380009, India); 
{  P.P. Kronberg$^{\dagger}$} (Dept. of Phys., Univ. Toronto, 60 St George Street, Toronto, ON M5S 1A7, Canada.); 
{  E. Kun} (Faculty of Physics and Astronomy, Ruhr$-$Universit\"at Bochum, Universit\"atsstra{\ss}e 150, D-44801 Bochum, Germany); 
{  A. Meli} (i) Dep. of Physics, North Carolina A\&T State Univ., Greensboro, NC 27411, USA; ii) Space sci. \& Techn. for Astrophys. Res. (STAR) Institute, Universit{\'e} de Li{\`e}ge, B-4000 Li{\`e}ge, Belgium);  
{  E.-S. Seo}, (Inst. for Physical Science and Technology, 4254 Stadium Dr, University of Maryland, College Park, MD  20742, USA). 
{  T. Stanev}, (Bartol Res. Inst. and Depart. of Phys. and Astron., Univ. of Delaware, Newark, DE 19716, USA) 

%
%



[1]{Department of Physics \& Astronomy, Washington State University, Pullman, WA 99164, USA}

[2]{Max Planck Institute for Radioastr., Auf dem H{\"u}gel 69, D-53121 Bonn, Germany}

[3]{Dept. of Phys. \& Astron., Univ. Alabama, Box 870324, Tuscaloosa, AL 35487-0324, USA}

[4]{Istituto Nazionale di Astrofisica $-$ Istituto di Astrofisica e Planetologia Spaziali, Via Fosso del Cavaliere 100, I-00133, Roma, Italy}

[5]{Astronomisches Institut, Ruhr$-$Universit\"at Bochum, Universit\"atsstra{\ss}e 150, D-44801 Bochum, Germany}

[6]{Centrum Astronomiczne im. Mikolaja Kopernika, Polish Academy of Science, Bartycka 18, PL-00-716 Warsaw, Poland}

[7]{Instituto de Astronom\'{i}a, Universidad Cat\'{o}lica del Norte, Avenida Angamos 0610, Antofagasta, Chile}

[8]{Institut f{\"u}r Kernphysik, Westf{\"a}lische Wilhelms-Universit{\"a}t M{\"u}nster, D-48149 M{\"u}nster, Germany}

[9]{Dept. of Phys. \& Astron., Univ. Alabama, Box 870324, Tuscaloosa, AL 35487-0324, USA}

[10]{Fak. Phys. \& Astron., Ruhr-Universit{\"a}t Bochum, Universit{\"a}tsstra{\ss}e 150, D-44801 Bochum, Germany}

[11]{International Center for Cosmology (ICC), Charusat University, Anand, GUJ 388421, India}

[12]{Dept. of Phys., Univ. Toronto, 60 St George Street, Toronto, ON M5S 1A7, Canada}

[13]{Fak. Phys. \& Astron., Ruhr-Universit{\"a}t Bochum, Universit{\"a}tsstra{\ss}e 150, D-44801 Bochum, Germany}

[14]{Dep. of Physics, North Carolina A\&T State Univ., Greensboro, NC 27411, USA} 

[15]{Space sci. \& Techn. for Astrophys. Res. (STAR) Institute, Universit{\'e} de Li{\`e}ge, B-4000 Li{\`e}ge, Belgium}

[16]{Inst. for Physical Science and Technology, 4254 Stadium Dr, University of Maryland, College Park, MD  20742, USA}

[17]{Bartol Res. Inst. and Depart. of Phys. and Astron., Univ. of Delaware, Newark, DE 19716, USA}


%
%

version Feb 13, 2024, revised Apr 26, 2024; revised again June 10, 2024
\end{centering}


\section{Abstract}
Supermassive black hole mergers with spin-flips accelerate energetic particles through their precessing relativistic jets, producing high energy neutrinos and finally gravitational waves. In star formation massive stars come in pairs, triplets and quadruplets, allowing second generation mergers of the remnants with discrepant spin directions. The Gravitational Wave (GW) data support such a scenario. Earlier we suggested that stellar mass black hole mergers (visible in M82) with an associated spin-flip analogously allow the acceleration of energetic particles, with ensuing high energy neutrinos and high energy photons, and finally producing gravitational waves. At cosmic distances only the gravitational waves and the neutrinos remain detectable. Here we generalize the argument to starburst and normal galaxies throughout their cosmic evolution, and show that these galaxies may dominate over Active Galactic Nuclei (AGN) in the flux of ultra-high energy particles observed at Earth. All these sources contribute to the cosmic neutrino background, as well as the gravitational wave background (they detected the lower frequencies). We outline a search strategy to find such episodic sources, which requires to include both luminosity and flux density.

Keywords: neutrinos, starburst galaxies, black hole mergers, gravitational waves, particle
acceleration 

\section{Introduction}

Searches for identification of the source of a given high energy neutrino, or gravitational wave event usually try to find both coincidences in direction on the sky, and some temporal coincidence, like excess emission at the same time. One of the best candidates for very high energy particle acceleration is the effect of relativistic precessing jets during the merger of two black holes. Such an event has been proposed to be identified in the starburst galaxy M82, by the action of the precession of a pair of powerful jets emanating from two stellar mass black holes (BHs) prior to their merger \cite{KBS85,Allen98,ASR18}. As we will show these jets match in their power the observed minimum of jet power of AGN \cite{Punsly11}, and so they can be quite efficient in producing ultra high energy cosmic ray (UHECR) particles, and as a consequence high energy neutrinos. In such a discussion it is important to note that energetic neutrinos might be highly boosted in the direction of the jet at the time of emission, and so additional selection effects operate, e.g. \cite{Kun17,Kun19,Kun21,Tjus22}.

We note that now both an extragalactic neutrino background (e.g. \cite{Jaroschewski23a,Jaroschewski23b}) and an extragalactic Gravitational Wave background have been discovered \cite{NanoGrav23a,NanoGrav23b,NanoGrav23c}, and a search strategy to identify sources is desirable.











\subsection{Binary star orbital angular momentum evolution}

An important question is whether it is possible that most or all stellar mass black holes can get born with near maximal rotation. There are two obvious mechanisms to get them to rotate fast: 

A first mechanism acts when the newly formed massive star has a rapidly rotating core, that remains in sufficiently high rotation until the star blows up as a SuperNova (SN), and  the BH forms \cite{Chieffi13,Limongi18,Limongi20}. That requires that angular momentum transport is small throughout the star, and also that the wind \cite{Seemann97} does not remove a significant quantity of rotational angular momentum throughout the life of the star. 

A second mechanism is plausible via a tidal locking,  since most massive stars are in binaries, triples or even quadruple systems. In the following we will work through the requirements for this path. This implies that binary stars in their evolution get close enough to actually achieve tidal locking \cite{Chini12,Chini13a,Chini13b}.

We will show that the removal of orbital angular momentum by the winds of the two stars is a key aspect.  

For didactic simplicity we consider two stars of equal mass $M$ at a distance of $2 \, r$ from each other orbiting in a circle with period $P$. Then the total orbital angular momentum is given by

\begin{equation}
J_{orb} \, = \, \pi^{-1/3} \, M^{5/3} \, G_N^{2/3} \, P^{1/3} \, , \nonumber
\end{equation}

\noindent where $G_N$ is Newton's constant of gravitation, and the radial scale $r$ can be connected to the other measures of the system by

\begin{equation}
r \, = \, \frac{1}{2 \, \pi^{2/3}} \, M^{1/3} \, G_N^{1/3} \, P^{2/3} \, . \nonumber
\end{equation}

It follows that the time changes are given by

\begin{equation}
\frac{\dot{J}_{orb}}{J_{orb}} \, = \, \left( \frac{5}{3} \, \frac{\dot{M}}{M} \, + \, \frac{1}{3} \, \frac{\dot{P}}{P} \right) \, , \nonumber
\end{equation}

\noindent and

\begin{equation}
\frac{\dot{r}}{r} \, = \, \left( \frac{1}{3} \, \frac{\dot{M}}{M} \, + \, \frac{2}{3} \, \frac{\dot{P}}{P} \right) \, . \nonumber
\end{equation}

The loss of orbital angular momentum by a wind is given by

\begin{equation}
{\dot{J}_{orb}} \, = \, 2 \, \dot{M} \, r \, {v}_{\phi} \, (1 \, + \, \varepsilon_{W,B}) \, , \nonumber
\end{equation}

\noindent where the term $\varepsilon_{W,B}$ describes the loss by the torque of the magnetic field (see \cite{Weber67}, their eq. (9)), and $v_{\phi}$ is the rotational velocity of the flow. Here we adopt the picture that the orbital radius acts as a lever arm.  It follows that the temporal evolution of the orbital radius is given by

\begin{equation}
\frac{\dot{r}}{r} \, = \, 2 \, \frac{\dot{M}}{M} \,\left(  \varepsilon_{W,B} \, - \, \frac{1}{2} \right) \, . \nonumber
\end{equation}

Next we need to put this into context: The angular momentum transport from both stars is given by  (\cite{Weber67}, their eqs. 8 and 9)

\begin{equation}
{\dot{J}_{orb}} \, = \, 2 \, \left( 4 \, \pi r^2 \rho \, {v}_r \, {v}_{\phi} \, r + B_r \, B_{\phi} \, r^3 \right) \, , \nonumber
\end{equation}

\noindent where ${v}_r$ is the radial velocity, and where the ratio of the second term and the first term gives $\varepsilon_{W,B}$. The first term corresponds to $2 \, \dot{M} \, r \, {v}_{\phi} $ above.

It follows that for mass loss, so for $\frac{\dot{M}}{M} \, < 0$, the orbital separation will increase for the case of no magnetic fields. But for $\varepsilon_{W,B} > 1/2$ the orbital separation will decrease. For equipartition in the wind $\varepsilon_{W,B} \, \simeq \, 1$. If the magnetic fields were really strong, so as to allow a lever arm even larger than the orbital radius $\varepsilon_{W,B} \, > \, 1$.

We conclude here that magnetic winds are the key for driving massive binary stars together, allowing locked in rotation. This gives rotation with the speed close to what had been assumed in the simulations of \cite{Limongi18,Limongi20}.

The ratio of the magnetic term to the flow term can be written as the inverse of two Alf{\'e}nic Mach-numbers:

\begin{equation}
M_{A, r} \, = \, \frac{{v}_r(r_*) \, \sqrt{4 \, \pi \, \rho(r_*)}}{B_r(r_*)}  \nonumber
\end{equation}

\noindent and 

\begin{equation}
M_{A, \phi} \, = \, \frac{{v}_{\phi}(r_*) \, \sqrt{4 \, \pi \, \rho(r_*)}}{B_{\phi}(r_*)} \, , \nonumber
\end{equation}

\noindent where $r_*$ is the radius, where density $\rho$, rotational velocity ${v}_{\phi}$,  and tangential magnetic field $B_{\phi}$ are evaluated. In the long distance limit, here ${v}_r$ goes to a constant, $\rho$ as $1/r^2$, $v_{\phi}$ as $1/r$, $B_r$ as $1/r^2$, and $B_{\phi}$ as $1/r$. It follows that $M_{A, r} \, \sim \, r$, and $M_{A, \phi} \, \sim \, 1/r$, so that the product $M_{A, r} \, M_{A, \phi} \, \sim \, 1$. Then 

\begin{equation}
\varepsilon_{W,B} \, = \, \frac{1}{M_{A, r} \, M_{A, \phi}} \, \sim \, 1 \, . \nonumber
\end{equation}

In a long list of OB stars compiled by Chini et al. from the literature 
\cite{Chini12,Chini13a,Chini13b} a typical orbital period is about 4 days, with quite a spread. The initial typical radius of these stars is about $10^{12.2} \, {\rm cm}$ almost independent of mass \cite{Chieffi13}, and so the inferred typical initial surface velocity is about 300 km/s, just the high velocity used in these calculations \cite{Limongi18}.  Surface magnetic fields are of order $10^3 \, {\rm  Gauss}$ \cite{Walder12}; however, the observational evidence suggests that some massive stars rotate more slowly with age rather than faster as argued here. That could happen, if the local angular momentum is maintained, so that the core rotates faster with time, and the outer parts of a star rotate more slowly with time. To get a quantitative estimate for $\varepsilon_{W,B}$ we have to adopt some further numbers: $v_r \, \simeq \, 2000 \, {\rm km/s}$, and for the magnetic field near the surface we adopt a low estimate of $B_r \, \simeq \, B_{\phi} \, \simeq \, 100 \, {\rm Gauss}$. For the mass loss we take $10^{-5} \, M_{\odot} \, {\rm yr^{-1}}$. This gives an estimate of $\varepsilon_{W,B} \, \simeq \, 1$. If the magnetic fields were any stronger, $\varepsilon_{W,B}$ would be larger, and then the orbital angular momentum loss would be yet stronger, allowing the two stars to get closer even faster. However, if the magnetic field were significantly weaker, this preponderance of the magnetic fields in removing orbital angular momentum would disappear, the two stars in a binary would move apart, rotate ever more slowly, and the spin of the resulting black hole might be far below maximal. The scant data \cite{Walder12} suggest that of the massive stars not all end up producing a rotating black hole, rotating near maximum; some do. The fraction of massive stars in binaries that do is unknown at present. 

However, there is the other option, mentioned at first above, that the cores of all massive stars are rotating fast right from the formation stage, allowing the surface to rotate much more slowly, and so deceiving any observer. This will be relevant also for all massive stars in binary systems, that do not tighten their orbit over time.

We will focus here on those stars that do produce a black hole rotating near the maximum allowed. All radio supernovae (RSNe) well observed seem to share a common property, namely that the product magnetic field times radius $(B \, \times \, r)$ has the same value in the wind \cite{ASR18,Biermann19}, comparing different radial scales $r$ and different RSNe in one galaxy, M82, as well as in different galaxies. Furthermore, we note that this quantity is consistent with what has been observed around the SMBH in the galaxy M87 \cite{EHT19}. Furthermore, the wind/jet power derived is consistent with the minimum jet power for radio-loud optically selected quasars \cite{Punsly11}. In many of the cases the central SMBH is believed to be near maximum rotation \cite{Daly19,EHT19}. Here we propose to explain this property also in stellar mass BHs as a result of the central black hole rotating near maximum at the beginning \cite{Chieffi13,Limongi18,Limongi20}, possibly reducing its angular momentum quite rapidly.

\subsection{Angular momentum of the black hole}

For all models the final predicted BH angular momentum is about 

\begin{equation}
J_{BH*} \, \geq \,  10^{51.1} \, {\left(\frac{M_{BH*}}{10 \, M_{\odot}}\right)}^2\, {\rm erg \, s} \, \simeq \,  10^{50.9} \, {\left(\frac{M_{BH*}}{10 \, M_{\odot}}\right)}^2\, {\rm erg \, s} \, = \, J_{BH,max} \, . \nonumber
\end{equation}

If there is an excess of angular momentum it has to be gotten rid off before a black hole can even form, even if near maximal rotation. There are several possibilities:

\begin{itemize}
\item{}  1st option:  A small initial BH mass near its spin limit grows, and sheds all excess angular momentum during growth through tidal gravitational torque or through magnetic torque. As massive star explosions are very clumpy, this might produce gravitational waves (GWs). No such waves have yet been detected.

\item{}  2nd option: the collapse first forms a binary BH (BBH), or a binary of a BH and a neutron star.  At each radius the angular momentum contained matches the limiting number allowed for that mass.  This implies, that we have maximal differential rotation, for BBHs near maximal individual spins are plausible  - individual spin-down has been shown to be slow \cite{King99}. This option would produce a high frequency GW event, and none has been seen as yet, that could be attributed to such a scenario for certain. On the other hand, three events have been seen with low mass partners (\cite{LIGO19,LIGO21,LIGO23,LIGO24}), which could be neutron stars or black holes. The sum of the two partners is consistent with the lowest mass BHs known. The aligned spin before the merger is consistent with zero in all three cases, which is expected in such a scenario. A bright SN showing the explosion of a very massive star is implied to accompany the final merger of the two fragments turned BH or neutron star.

\item{}  3rd option: There is a burst of ejected excess angular momentum and energy via magnetic fields: this is akin to a proposal by \cite{GBK70}, and many papers later, such as \cite{GBK08}. He proposed that this is the mechanism to explode massive stars, to make a SN.

\item{}  4th option: A collapse into a Kerr geometry, with $(J_{BH} \, c)/(M_{BH}^2 \, G_N) \, > \, 1$, is allowed \cite{Joshi20}. This is still an astrophysical BH (i.e. lot of mass compacted in small volume, with no event horizon). There are powerful mechanisms, how such (a naked singularity) configuration very rapidly gives away angular momentum, and settles to a rotating BH with a horizon. Here, one gets the required burst-like energy also from high angular momentum decay. 

\end{itemize}

All options listed here lead to a BH in near maximal rotation, a state which may last only a short time. So we will assume near maximal rotation for now, and revisit these arguments later again. If there is no excess to start with, the angular momentum can still be very close to maximal according to the simulations of \cite{Limongi20}.

\section{Black hole mergers, supernovae and other episodic events}

Here we focus on stellar mass BH mergers, as one example of a short injection of energetic particles, recognizable via the cone of precessing jets, that clean out the ISM (e.g., source 41.9+58 in the starburst galaxy M82, \cite{KBS85,Allen98,ASR18}).

\subsection{Source 41.9+58, a second generation stellar mass black hole merger?}


The compact radio source 41.9+58 sits at the apex of a triangular region without radio emission opening south, with a less regular region without radio emission to the north \cite{KBS85}; a detailed image is in \cite{ASR18}. The difference can be understood as the result of projection effects, since the disk of M82 is slightly tilted relative to the line of sight. This can be interpreted as the action of a pair of two-sided precessing jets emanating from two coalescing active rotating black holes of stellar mass \cite{KBS85,ASR18}. As most massive stars sit in stellar binary systems, triples and often even quadruples, each close binary will interact such that their spins can be expected to align, while distant binaries resulting from two first generation mergers of two stars or black holes each can be expected to yield very different spin directions.  Magnetic winds help bring two stars or two black holes together by removing orbital angular momentum. The large cone of precession results in the case, that the two black holes initially have vastly different spin directions, and the black holes slowly align their spin directions before their actual merger \cite{Gergely09}.  This {to\-po\-logy} is inconsistent with an explosion in a stratified atmosphere, since that always leads to a stem-like outflow (extensive literature is given in \cite{ASR18}). Such stem-like outflows are in fact seen as filaments above and below the disk of M82 \cite{ASR18}.  

Could there be other such features hidden in the radio map of the inner region of M82 \cite{KBS85}? If a large proper motion were to be allowed, then there are a number of possibilities to allow an interpretation of another such double-cone feature, with source 44.0+59.5 a speculative option.

So the detection of one such source out of 43 yields a very uncertain estimate of their rate of one per 2,500 years in the starburst galaxy M82 \cite{ASR18}. M82 has a Far Infra-Red (FIR) dominated luminosity of about $10^{10.6} \, L_{\odot}$ \cite{KBS85}, and so that rate can be estimated to be correspondingly higher for a higher FIR luminosity.

\subsection{Fraction of mergers among massive stars}


In M82 we observe 43 compact sources \cite{KBS85}, probably all of which are explosions of Blue Super Giant (BSG) stars, since the winds of Red Super Giant (RSG) stars do not provide enough ram pressure to allow the quick formation of Radio Super Novae (RSN) of the size as observed, of a few parsec \cite{KBS85,Allen98,Allen99,ASR18,Biermann19}. We find a single source, 41.9+58, which appears to be fully consistent with a second generation BH merger.  The FIR luminosity of M82 can be interpreted as a measure of the star formation rate. The SN rate for massive stars (i.e. all above a Zero Age Main Sequence (ZAMS) mass of about $10 \, M_{\odot}$) can be estimated to be within the range of 1 per 1.5 years and 1 per 5 years \cite{KBS85,ASR18}, and so the rate of such second generation mergers can be very crudely estimated to 1 in 1000 of massive stars, with an error range of probably at least an order of magnitude.

\subsection{Rate of mergers}


Using a scaling with FIR luminosity yields a maximal rate of $10^{12}/10^{10.6} \, \times \, 1/2,500$ per year, so about 1 in about 100 years at most. This is again an order of magnitude estimate only.

What is exactly the scenario of energetic particle injection? Powerful plasma jets precess and therefore continuously encounter new material to accelerate to ultra high energies. This new material is fed to the central region of the starburst galaxy by friction in the Interstellar Medium (ISM) \cite{Toomre72,Wang00}, in the model to consider any gaseous galaxy akin to an accretion disk \cite{Luest52}. Starburst galaxies often involve the merger of two galaxies, stirring up their ISM \cite{Toomre72}.

Then the next question is the length of time of the active episode: For that we use column 2 of Table 2 in \cite{Gergely09}, and so the initial inspiral rate, so scaling the expression in the last line, for the angle change, to $10 \, M_{\odot}$ and an equal mass binary gives a time scale of 5 years, so still a small fraction of 100 years. This implies that in our model, the injection of energetic particles due to the inspiral motion of active black holes is taken to last of order 5 years (this time scale scales linearly with the mass). Therefore, the precessing motion makes the injection of new particles ready to be accelerated so very much more efficient than a non-moving jet. Thereafter, when the merged BH starts another pair of jets, injection of energetic particles continues, but at a much lower rate, since the precessing motion has ceased, and so the encounter with new material is reduced.

This time scale is based on the initial stage, when gravitational wave emission becomes the dominant means to remove orbital angular momentum \cite{Gergely09}. We have proposed above that magnetic stellar winds, using the angular momentum lever arm of the orbital radius, remove sufficient orbital angular momentum to get the system to this point.

This time scale is short compared with the time scale between such events, estimated above at order 100 years for the most luminous starburst galaxies, and longer for starburst galaxies of lower FIR luminosity. Therefore it appears possible but fairly unlikely that any starburst galaxy will experience many such activity episodes at the same time.

\subsection{Episodic activity and corresponding energies}


Therefore, for a starburst galaxy of a FIR luminosity $L_{FIR}$ other than the maximum of $10^{12} \, L_{\odot}$ the time-scale between such episodes of injection is then correspondingly longer than 100 years, so is of order $100 \, {\rm yrs} \, \{10^{12} \, L_{\odot}/L_{FIR}\}$.

This implies that in any given flux density interval of a sample those galaxies contribute the most, that have the highest FIR luminosity, so are at the highest redshift. They have the highest probability to be in an active stage right now (in the observer frame), as compared to other galaxies at the same flux density but lower redshift. This is a key step in the argument proposed.

If BH spin energy drives powerful jets, it implies that the rotational energy is available: That implies that for a final mass of 10 $M_{\odot}$ we have $\{\sqrt{2} - 1 \} \; M_{BH} \, c^2$ maximally available to drive a magnetic jet, replete with energetic particles. For a $10 \, M_{\odot}$ final mass this is some fraction of $10^{54.9} \, M_{BH, 1} \, {\rm erg}$ times an inefficiency factor that estimates what fraction of this energy goes into energetic particles. Allowing $1/3$ this gives $10^{54.4} \, M_{BH, 1} \, {\rm erg}$. Counting at first only the second generation mergers happening every 2,500 years (note the uncertainty in this number!) it implies that potentially we have a power input of $10^{43.8} \, M_{BH, 1} \, {\rm erg/s}$, noting that this involves two such black holes. The minimum power required in M82 to clean out the ISM \cite{ASR18} can be estimated as follows: First of all the $P \, dV$ work can be estimated by using the numbers in \cite{KBS85}: The volume is a cone of about 50 pc baseline radius, and about 30 pc height, giving a volume of about $10^5 \, {\rm pc^3}$; the pressure can be estimated also from \cite{KBS85} as about 4 times the magnetic field pressure (magnetic field, energetic particles, and thermal particles giving a pressure equal or larger than magnetic fields and energetic particles combined), so using a magnetic field strength of $10^{-3.7} \, {\rm Gauss}$ this gives a pressure of $10^{-8.8} \, {\rm dyn}$. The total $P \, dV$ work is then $10^{51.7} \, {\rm erg}$. Since we are referring to the sweeping action of the precession cone, the time scale has to be that for changing the angle: as derived above, this yields 5 yrs for this time scale, assuming for reference again $10 \, M_{\odot}$, and so the associated power flow has to be of order $10^{43.5} \, {\rm erg/s}$ for two jets, so $10^{43.2} \, {\rm erg/s}$ for one jet. This is in fact consistent with the power flow derived from the quantity $(B \times r) \, = \, 10^{16.0 \pm 0.12} \, {\rm Gauss \times cm}$ observed for the common magnetic field in young RSN (\cite{ASR18,Biermann19}; based on \cite{KBS85,Allen98,Allen99}), using the approach of \cite{Falcke95}): This yields $10^{42.8} \, {\rm erg/s}$, easily within the errors of such a comparison. This derivation is independent of BH mass, as the consistency with the minimum power in radio quasars \cite{Punsly11}, and with the magnetic field in the M87 radio core \cite{EHT19}. This is a consistency check on the power flow in the precessing jets. At this point we can derive the time scale of angular momentum loss and energy loss: This can be determined by dividing the maximally available energy of $10^{54.9} \, M_{BH, 1}$ by this luminosity derived here of $10^{42.8} \, {\rm erg/s}$, which gives $10^{12.1} \, {\rm s} \, M_{BH, 1}$; here $M_{BH, 1}$ is the mass of the black hole in units of ten Solar masses. On this time scale a maximally rotating black hole loses angular momentum and energy, at the minimum. This shows that for a BH mass of $10^{6.6} \, M_{\odot}$ we reach the lifetime of the universe. Curiously, this happens to be the mass of the SMBH in our Galactic center, for which its rotation state is not yet known \cite{EHT22}. The power derived here is slightly lower than the power output derived at the beginning, of $10^{43.5} \, {\rm erg/s}$ for one BH; a simple interpretation may be that there are channels other than the magnetic jet itself to use up the rotational energy of the black hole, e.g., via the Penrose process \cite{Penrose71}, or even simpler that the life-time of the high spin of the black holes is just longer than the merging time scale; since many black holes get a kick at formation, they leave the galaxy, and the detections of RSNe in M82 may be limited by these objects just flying out. If this is the correct understanding, then all these rotating black holes are flying through the region around galaxies like M82, and lose most of the rotational energy out there.

If this rotational energy of a rotating black hole is ejected via magnetic fields and energetic particles, in a relativistic wind or jet, could their contribution to energetic particles in intergalactic space surpass the contribution from Super Massive Black Holes (SMBHs)? The combined useable rotational energy of all these stellar mass BHs can be estimated for our Galaxy, following the summary of the data in \cite{ASR18}, based on \cite{Diehl06}, using $10 \, M_{\odot}$ again as reference for simplicity, as about $10^{62.4} \, {\rm erg}$, to be compared with the maximal useable rotation energy of our Galactic Center BH, assuming that it ever achieved this, as $10^{60.5} \, {\rm erg}$. This all depends on interpreting these stellar mass BHs beginning with a near maximal rotation state, as suggested by the commonality of the magnetic field in RSNe, and the simulations by \cite{Chieffi13}, and \cite{Limongi18}; the argument has been given above in detail. If these stellar mass BHs also produce relativistic jets, the maximum energy particles may reach well beyond the ankle in the CR spectrum. By these same magnetic fields they lose their rotation quite fast, in about $10^{4.2} \, {\rm yrs}$ for a $10 \, M_{\odot}$ BH (above we derived a similar number, $10^{4.6} \, {\rm yrs}$, using energy output). Summed over the lifetime of our Galaxy, this corresponds to a power input of $10^{44.7} \, {\rm erg/s}$ outside our Galaxy, today it is a factor of order 2 less, so about $10^{44.4} \, {\rm erg/s}$. The Galactic CRs require an input of order $10^{41.0} \, {\rm erg/s}$ \cite{Gaisser13}, which gives an efficiency of about $10^{-3.5}$ for CR injection inside the CR disk.  As the typical galaxy density is of order $10^{-2} \, {\rm Mpc^{-3}}$ \cite{Lagache03}, and an order of magnitude lower at the FIR luminosity of our Galaxy, using this efficiency, it yields a crude estimate of $10^{39.0} \, {\rm erg/s \, Mpc^{-3}}$.  This can also be checked directly with the density of SMBHs (e.g., \cite{Caramete10}) of $10^{5.5 \pm 0.4} \, {\rm M_{\odot} \, Mpc^{-3}}$, which corresponds to a maximally useable CR energy flow of $10^{38.2} \, {\rm erg/s \, Mpc^{-3}}$, which is slightly less than the possible contribution from massive star BHs, but consistent within the uncertainties. On the other hand, SMBHs can accrete and power outflows also at the Eddington limit, and that yields a very much higher possible power inputs for a short time: Using the same densities of SMBHs \cite{Caramete10} and a time fraction of order $10^{-2}$ for high activity yields then about $10^{41.6} \, {\rm erg/s \, Mpc^{-3}}$, still below the purely spin-down based stellar mass BHs power input, derived above, of $10^{44.4} \, {\rm erg/s}$.  This can be compared with the average UHECR energy input worked out by, e.g., \cite{Waxman95} of $10^{37.1} \, {\rm Mpc^{-3} \, erg/s}$. The possible contribution from massive stars exceeds the AGN UHECR contribution. So massive star BHs may make a substantial contribution to UHECRs, in the case of initially high rotation, and relativistic jets, as implied by the M82 observations. This is fully consistent with new Auger results \cite{Auger24}. 

Finally, there is another consequence of this minimum loss time for angular momentum. In a star cluster of massive stars these stars also lose orbital angular momentum via their magnetic winds, setting up a merger of massive stars to form a super-massive star (see, e.g., \cite{Spitzer69,Sanders70,Wang00}), which in turn may quickly form a SMBH of a mass close to the GC SMBH (\cite{Appenzeller72} focus on the explosion only). A fortiori this also works for the merger of stellar mass BHs. This process can speed along the early formation of SMBHs, as observed by JWST (e.g., \cite{Uebler23}).

\subsection{Other sources of episodic activity}


The classical episodic events to inject energetic particles are super-nova (SN) explosions first of all (e.g. \cite{Cox72}). But normal SN explosions running through a former stellar wind give a maximal particle energy of about $10^{17.5} \, Z$ eV, so reach the ankle, but certainly do not go beyond \cite{ASR18}. The reason is that in such RSNe the magnetic field in terms of $B \times r$ is observed to be always close to the measure $10^{16.0 \pm 0.12} \, {\rm Gauss \times cm}$ \cite{ASR18,Biermann19}, as extensively discussed above. In the well observed sources there is not clearly a large tail of this quantity on either side of this specific number. However, the selection effects could be large in such a tally. However, if these rotating black holes were to initiate a relativistic jet \cite{Mirabel99}, the particle energies accelerated could go much higher.

Other episodic sources are binary star systems with one BH, pulsars and pulsar winds, white dwarf SNe (SN I a), active neutron stars in binary systems, as well as neutron star mergers.

It is important to add, that a further source of episodic acceleration can be due to electric discharges \cite{Gopal24}: Winds and jets patterned after the Parker wind \cite{Parker58} carry an electric current. When the power varies with time, then the electric current changes. This change builds up electric charges and fields following Maxwell's equations \cite{Gopal24}, here the equation of continuity for electric currents which is contained in Maxwell's equations. These electric fields can discharge violently and produce acceleration of particles (see for the possibility of an electric discharge close to the central BH \cite{Aleksic14}); in the limit of strong electric fields this discharge acceleration produces a 1D momentum $p$ spectrum of $p^{-2}$, quickly scattered to a 3D $p^{-4}$ spectrum. This spectrum has been recognized in radio emission in radio filaments that may have undergone an electric discharge \cite{Gopal24}, both Galactic and extra-galactic. The magnetic irregularity spectrum excited by this steep particle spectrum also contributes to a good fit to the newest AMS data for Fe energetic particles \cite{Allen24}, and presumably also for other primary elements like He, C and O, with the difference, that He, C, and O have spallation additions from higher elements and not only spallation losses like Fe.

\subsection{Probability}


The probability that a given starburst galaxy is ejecting for instance high energy neutrinos right now (in the observer frame) runs with the FIR luminosity in our proposed model. Therefore comparing all sources at some given flux density those at the highest luminosity, therefore highest redshift, have the highest probability to contribute. As shown above, there is probably no case, where multiple activity contributors are relevant at the same time. 

To go beyond identifying most probable sources, say by working out the total neutrino background, we go one step further: Once the sources are summed weighted by probability we follow by adding all different flux density levels (cf. \cite{Caramete16}).
 


Clearly, a merger of two stellar mass black holes with the associate precession of jets aligning each with the spin of a black hole, is likely to accelerate particles to high energies, so that interaction takes place, and neutrinos are emitted. At the very end of this stage, the two black holes will merge, and emit a burst of gravitational waves. It is important to note that due to boosting the selection effects governing the detection of neutrinos and a burst of GWs are very different.  So the detection of both due to the same episode of a source at about the same time is unlikely.



The main aspect in the analysis is that at any given flux density the sources with the  highest intrinsic luminosity, so highest redshift, have the highest probability to contribute. This would be the same conclusion for the other possibilities of episodic injection of energetic particles, such as SNe. However, if the energetic particles are stored and not ejected via the open precession cone, then the line of reasoning is valid only if most of the interaction happens right at the start, as has been argued already \cite{Stanev93,Biermann01,ASR18,Allen24}.

\subsection{Analogy of super-massive black hole mergers?}


This approach may be useful as well for Active Galactic Nuclei (AGN) with central Supermassive BHs (SMBHs), since their activity is also episodic. Assuming that relativistic boosting is not stronger for minimum power AGN-BHs at near maximal rotation, then looking for the highest luminosity within a given flux density interval should also give a higher probability for the source to give either neutrinos or gravitational waves. For many AGN the FIR range has the highest probability to actually be strongly influenced by thermal dust emission (e.g. \cite{Chini89a,Chini89b}), powered by the activity of the central SMBH. The flat spectrum AGN S5 1803+784 is a famous counter-example with its FIR emission in line with a flat spectrum extrapolation from 5 GHz \cite{Gregorini84}.

So we tentatively propose for AGN-BHs a similar observing strategy as for starburst galaxies, with a focus on the FIR: Take spectra of all sources in the plausible search window on the sky, including the FIR continuum. Then select a flux density interval, and pick a sample of the highest luminosity sources among them. Try to verify whether any of them could be the source; if unsuccessful, pick another flux density interval, and repeat the exercise.

So a similar approach might be useful to test, to select at any given flux density the highest luminosity sources, with two approaches, first to go for the FIR dust emission, and second for the FIR flat spectrum extension. 

\subsection{An observational strategy}


Consider the detection of a gravitational wave event, or alternatively the detection of a high energy neutrino event, likely to be of astronomical origin. Then first an area needs to be identified that may contain the galaxy with the source. Thereafter take a spectral map of this area, which shows the approximate redshift for all sources. 

Proceed as follows:

i) Rank all candidate sources in FIR flux density.

ii) Start with the galaxy at the highest flux density, and then define the (index j, here $j \, = \, 1$), the first sample (index $i$) by

\begin{equation}
\Sigma  L_{FIR, j, i} \, > \, L_{FIR, M82} \, \frac{\tau_{3.4}}{\tau_{ep}}\, , \nonumber
\end{equation}

\noindent where $\tau_{3.4}$ the repetition time scale is, in our BH merger approach $2,500 \, {\rm yrs}$, and $\tau_{ep}$ the length of the UHECR injection, in our approach the length of the time, during which the jets precess, 5 yrs. Thus, in this sample, there is a $\simeq \, 100$ percent expectation, that some galaxy is in an active phase of an episode. The size of the sample is one parameter. The flux density interval chosen needs to be large enough, so that subsequent intervals do not overlap in combined probability of identification.

iii) Then rank within the sample all sources by FIR luminosity. The galaxy with the highest luminosity has the highest probability to be the real source.

iv) Repeat, using the next group of galaxies (index $j$), and use the same size of the sample, by adjusting the next flux density boundary; for a Euclidean universe one choice could be stepping flux densities by a factor of $2^{-2/3}$, so that we get equal and large numbers at each step.

Check the candidates in the set for any sign of activity that may relate to the event chosen, like visible variability. Considering the observations of M82, a sign would be if a compact source changes structure or spectrum as 41.9+58 did.  If there is no such sign, pick the next set of lower flux density, and repeat the exercise. Iterate the procedure, until successful, or until the observations run out of sensitivity.

Clearly, this needs a learning experience, different for every class of sources identified.  We chose this model to emphasize the possibility for the maximal energy to go beyond the ankle, near $10^{18} \, {\rm eV}$, and do so with a high rat of injection into the acceleration process.  Our model as proposed can be justified only for starburst galaxies, and it remains to be tested whether an analogous approach might be helpful also for AGN.

\section{Conclusion}

We propose a model and a two-step strategy to identify sources for either high energy neutrinos, or gravitational waves, based on the concept that that their production and emission from starburst galaxies is episodic, with the probability that the galaxy contains an emitter currently active in the observer frame running with the FIR luminosity, and the probability that we actually detect the emission running with the flux density. An analogous approach for AGN might be similar, but remains to be developed, justified and tested.

\end{document}